\documentclass{interact}

\usepackage{epstopdf}
\usepackage{subfigure}
\usepackage{xurl}
\usepackage{hyperref}
\usepackage{color}
\usepackage{widetable}
\usepackage{longtable}

\usepackage{natbib}
\bibpunct[, ]{(}{)}{;}{a}{}{,}
\providecommand{\bibfont}{}
\renewcommand{\bibfont}{\fontsize{10}{12}\selectfont}

\theoremstyle{plain}

\theoremstyle{definition}

\theoremstyle{remark}

\begin{document}

\articletype{ARTICLE TEMPLATE}

\title{Crypto Price Similarity: An Investigation Using Pearson Correlation and Dynamic Time Warping}

\author{
\name{Onur Batin DO\u{G}AN\textsuperscript{a} and Fatma Sevin\c{c} KURNAZ\textsuperscript{a,b}\thanks{CONTACT F.~S. Kurnaz Email: fskurnaz@yildiz.edu.tr; fatma.kurnaz@case.edu}}
\affil{\textsuperscript{a}Yildiz Technical University, Istanbul, Turkiye; \textsuperscript{b}Case Western Reserve University, Ohio, USA}
}

\maketitle

\begin{abstract}
The cryptocurrency market's volatility and complex price dynamics challenge portfolio diversification and risk management. This study examines price similarity across 347 cryptocurrencies from March 2020 to April 2024 using daily OHLCV data from the Binance API. Daily dynamics are summarized via a signed, capped intraday price-range metric, and pair similarity is assessed using the Pearson cross-correlation coefficient, capturing linear co-movement at a given lag, and Dynamic Time Warping (DTW) distance, capturing shape similarity independent of lag. Data are segmented into three market phases: an uptrend, a decline, and a subsequent uptrend. For each period, the 10 pairs with highest cross-correlation and lowest DTW distance are identified. BNB-CAKE, XLM-XRP, and MANA-SAND show the strongest cross-correlations across the three periods, with coefficients of 0.78 or higher and maximum correlations at lag 0, indicating synchronized movements. Three to four of the ten lowest-DTW pairs coincide with the cross-correlation top 10 per period, showing the measures are related but not interchangeable. Two same-platform fan tokens rank among the most DTW-similar pairs despite weak cross-correlation, while a euro-pegged asset and gold-backed token show the lowest DTW distance, driven by low volatility rather than shared dynamics. Results demonstrate substantial, measure-dependent interconnectedness with implications for diversification strategies.
\end{abstract}


\section{Introduction}
\label{sec:intro}

The cryptocurrency market has become one of the fastest-growing financial markets in the world \citep{almeida2022portfolio}, increasing from \$1 trillion to \$3 trillion during Q4 2024 \citep{grayscale2025}. This remarkable growth can be traced back to the introduction of Bitcoin, which is widely recognized as the world's first digital and decentralized currency, created in 2008 under the pseudonym \citep{nakamoto2008bitcoin}. Unlike traditional fiat currencies, which are governed by central authorities, Bitcoin operates on a decentralized network, allowing traders and participants to conduct transactions without intermediaries. Its innovative structure is built on blockchain technology, which is regarded as a significant development for ensuring security and preserving privacy \citep{gadekallu2022}, enabling secure and transparent transactions recorded on an open-access ledger. Blockchain technology has also been viewed as a significant innovation, with some cryptocurrencies potentially serving as assets that reflect a stake in the future development of this technology \citep{liu2022common}. Following the introduction of Bitcoin, thousands of cryptocurrency-related businesses have emerged, providing various services and innovations across multiple sectors \citep{Dash2023}. These developments have demonstrated the extensive applications of blockchain technology beyond digital currencies. The widespread adoption of cryptocurrencies and blockchain technology has the potential to reshape several sectors, including finance \citep{Shalini2023}, healthcare \citep{Saxena2023}, and transportation \citep{Dadsena2023}.

In addition to the impact of blockchain technology, the world's monetary landscape has also begun to change in recent years \citep{ALLEN2022102625}, driven by the emergence and rapid expansion of cryptocurrencies. Cryptocurrencies have increasingly been considered popular investment assets among investors \citep{LI2021101389}. The launch of spot ETFs offers institutional investors simplified access, improved alignment with spot prices, and stronger adherence to regulatory standards. This development is expected to attract substantial institutional participation, potentially fostering greater market stability and efficiency \citep{krause2024rise}. Nevertheless, several studies have raised concerns about market efficiency and volatility \citep{almeida2024cryptocurrency}. Evidence supporting inefficiency in cryptocurrency markets has been documented in several studies \citep{AGGARWAL2020100335,akyildirim2021prediction,caporale2018persistence,grobys2020predicting,sapkota2021asset,takaishi2018taylor,vidal2019weak}. Such inefficiencies may contribute to strong herding behavior, in which investors imitate the behavior of others rather than relying on independent analysis strategies \citep{BOURI2019216}. Inefficiencies may also lead to diminished diversification benefits as cross-asset correlations increase during periods of market stress \citep{CORBET201828}. These interconnected effects demonstrate the increased risks associated with investing in a market characterized by varying degrees of efficiency. On the other hand, there is also evidence suggesting greater market efficiency. For instance, the cryptocurrency market has been found to be more efficient than expected and to exhibit a strong low-volatility premium \citep{burggraf2021fears}, while other evidence suggests that it has become increasingly efficient over time \citep{ALVAREZRAMIREZ2021109997}. An increase in cryptocurrency market efficiency has also been associated with a decrease in average price delays \citep{KOCHLING201939}. Similar complexity is observed in the volatility characteristics of the cryptocurrency market. The dynamics of the cryptocurrency market exhibit two states, stable and volatile, which vary across cryptocurrencies in terms of volatility, average return, and inter-state dynamics \citep{bejaoui2019market}. The inherent volatility of cryptocurrencies also exposes investors to uncertainty and various risks \citep{liu2022common}. These discussions emphasize the need for a detailed understanding of market dynamics across different market phases.

Cryptocurrency market behavior and interconnectedness represent additional important dimensions of these dynamics. The cryptocurrency market exhibits different degrees of long-range dependence, which reflects the extent to which previous price changes affect future values over long time horizons. Cryptocurrency price dynamics may also follow a range of stochastic processes that describe the random and probabilistic characteristics of price movements \citep{BARIVIERA2021101649}. Many widely traded cryptocurrencies tend to exhibit monofractal behavior, meaning that their dynamics can be described in terms of relatively homogeneous and constant scaling properties. In contrast, other cryptocurrencies exhibit strong multifractality, whereby scaling properties vary and become more complex, reflecting heterogeneous and irregular market dynamics \citep{BARIVIERA2021101649}. Additionally, dynamic conditional correlations among cryptocurrencies suggest that their relationships are sensitive to market events and speculative activity \citep{kostika2020dynamic}. The interconnectedness of the cryptocurrency market further highlights its complex and dynamic nature. Significant spillover effects have been documented across the market \citep{TIWARI2020101083}, indicating a high degree of interconnectedness \citep{CORBET201828}. However, this interconnectedness is not static and varies over time \citep{ASLANIDIS2019130}. Total dynamic connectedness has been reported to fluctuate between 25\% and 75\%, with higher levels of connectedness during periods of greater uncertainty and lower levels when uncertainty is comparatively low \citep{ANTONAKAKIS201937}. In addition, bad contagion, referring to the spread of adverse events or shocks across assets, has been observed throughout the cryptocurrency market \citep{SHAHZAD2021101797}.

Various methodological techniques have been employed to analyze relationships between cryptocurrencies. Ji et al. \citep{JI2019257} utilized dynamic conditional correlation models to assess interrelations among major cryptocurrencies and concluded that correlations tend to increase during unstable market conditions. Complex correlation networks among cryptocurrencies have also been investigated using wavelet coherence analysis of return and volatility co-movements \citep{QIAO2020101541}. Wavelet methods have further been used to investigate interdependencies among major cryptocurrencies and a cryptocurrency volatility index, revealing considerable and persistent positive correlations across various investment time scales \citep{Agyei2022}. Furthermore, dynamic equicorrelations among cryptocurrency markets have been investigated to illustrate the temporal evolution of interconnectivity patterns \citep{DEMIRALAY2021524}. Qureshi et al. \citep{QURESHI2020125077} studied the interdependence of cryptocurrency markets using time-frequency approaches and identified considerable differences in correlation structures across various time horizons.

While prior research has examined various aspects of cryptocurrency market behavior, including efficiency, volatility, and interconnectedness, there remains room for further investigation of correlation dynamics across distinct market phases. First, much of the existing research focuses on a relatively small selection of cryptocurrencies, typically those with the largest market capitalizations, which may overlook important dynamics in the broader market. Second, many studies examine relatively limited time periods, which may not capture a broad range of market conditions and associated changes in correlation structures. Third, the role of time lags in cryptocurrency correlations remains of interest, particularly given evidence that information may be incorporated into cryptocurrency prices at different speeds. This study addresses these issues through the following research question: ``How does price similarity among 347 cryptocurrencies vary over a four-year period across different market phases?'' The main objective is to examine changes in these correlations across market phases and to provide insights relevant to portfolio construction and risk management. Specifically, this study analyzes time-lagged Pearson cross-correlations, complemented by Dynamic Time Warping distance, using multi-year data obtained from Binance, one of the world's leading cryptocurrency exchanges. The use of a broad set of cryptocurrencies across multiple market phases provides an opportunity to examine how correlation patterns vary under different market conditions. The resulting analysis contributes to the empirical understanding of cryptocurrency market dynamics and provides practical insights into the changing nature of relationships among cryptocurrency assets. In particular, examining these relationships across different market phases may help inform diversification strategies that account for changes in cryptocurrency correlation structures.

The remainder of the paper is organized as follows. Section \ref{sec:methods} describes the methodology, including the data collection procedures, selection criteria for cryptocurrency pairs, and the cross-correlation and Dynamic Time Warping frameworks used to assess similarity between cryptocurrency pairs. Section \ref{sec:results} presents and interprets the main findings, including the estimated cross-correlation coefficients, DTW distances, and corresponding data visualizations. Finally, Section \ref{sec:conclusion} summarizes the main findings, discusses their practical implications for investment strategies, and outlines the limitations of the study and directions for future research.

\section{Materials and Methods}
\label{sec:methods}

This study examines the degree of similarity in price movements among different cryptocurrencies using Pearson correlation analysis, complemented by Dynamic Time Warping (DTW) as a secondary, shape-based similarity measure. The Pearson correlation coefficient, also referred to as the Pearson product-moment correlation coefficient, is a statistical measure of the linear relationship between two variables \citep{Hair19}. As described by Hair et al.\ \citep{Hair19}, the coefficient measures the extent to which two variables move together and ranges from $-1$ (a perfect negative linear relationship) to $1$ (a perfect positive linear relationship), while a value of $0$ indicates the absence of a linear relationship. Pearson correlation was selected for this study because it provides a straightforward and interpretable measure of both the direction and strength of a linear relationship \citep{meissner2015}. Because Pearson-based cross-correlation is inherently limited to rigid, integer-day lag shifts and to linear co-movement, it was supplemented with DTW, which allows for elastic alignment between two series and can therefore detect shape similarity even when price dynamics are not perfectly synchronized in time. Together, the two measures provide a more complete picture of similarity between cryptocurrency price dynamics than either would provide alone.

\subsection{Data Acquisition}

For this study, daily OHLCV (Open, High, Low, Close, Volume) data were obtained through the Binance API, with all prices expressed in USDT. Cryptocurrencies were included in the initial sample if they were listed on Binance with an available USDT-quoted daily trading history. 
After removing cryptocurrencies delisted before the end of the observation window (Section~2.2), the retained daily series span March 1, 2020, through April 15, 2024, although the last trading day with observed data for any cryptocurrency was April 8, 2024.
The data were divided into three distinct market phases: an increasing trend from March 12, 2020, through November 10, 2021; a decreasing trend from November 11, 2021, through October 12, 2023; and a subsequent increasing trend from October 13, 2023, through April 15, 2024. OHLCV data summarize the price movements of financial instruments over specified periods and include the opening price at the beginning of each period, the highest and lowest prices reached during the period, the closing price at the end of the period, and the volume of assets traded. OHLCV data are commonly used in financial market research to examine price trajectories, volatility, and other important market indicators \citep{mann2017}. The Binance API was selected as the data source because Binance is one of the world's leading cryptocurrency exchanges, and its API enables programmatic access to historical market data.

\subsection{Data Cleaning \& Transformation}

The cryptocurrency price data underwent preprocessing to ensure their suitability for subsequent analysis. The data were inspected for missing values, which may occur for several reasons, including exchange downtime, API limitations, and the delisting of cryptocurrencies from Binance. Cryptocurrencies whose last available trading date did not coincide with the latest date present in the overall dataset were treated as delisted and removed entirely, so as to preserve time-series integrity for the remaining coins. Following this procedure, \textbf{347} cryptocurrencies were retained for the correlation analysis reported in Section~3. Within each of the three market phases, a cryptocurrency was further required to have no more than 80\% missing daily closing prices during that phase in order to be included in the corresponding phase-specific analysis; this yielded \textbf{198} cryptocurrencies for Period 1, \textbf{325} for Period 2, and \textbf{347} for Period 3.

\subsection{Calculations}

Rather than close-to-close price changes, daily price dynamics were summarized using a directional intraday range metric, and this metric was used to examine similarities in cryptocurrency price movements. For a given cryptocurrency on a given day, with $O$, $H$, $L$, and $C$ denoting the open, high, low, and close prices, respectively, the metric is defined as

\[
R = \min\!\left(\frac{H-L}{L},\ 2\right), \qquad
X = \begin{cases} R, & \text{if } C > O \\ -R, & \text{if } C \le O \end{cases}
\]

so that $X$ reflects the magnitude of the intraday high-low price range, capped at $2$, signed according to whether the day closed above or below its opening price. In practice, $|X|$ exceeded $1$ for approximately $0.16\%$ of daily observations across the full dataset, reflecting occasional large intraday swings, particularly for lower-liquidity coins.
Because cross-correlation analysis is conventionally interpreted under weak-stationarity considerations, the analysis was conducted on the bounded daily $X$ series rather than on cryptocurrency price levels (Venables and Ripley 2002).
Strict stationarity requires the statistical properties of a time series to remain unchanged over time and may be too restrictive in many practical applications. A weaker condition, known as second-order stationarity, requires a constant mean and variance and a covariance structure that depends only on the time lag between observations rather than on the specific time points. 

Suppose that a time series $X_t$ extends over time but is observed only at $t = 1, \dots, n$. A time series is said to be second-order stationary if the following conditions are satisfied \citep{Venables02}:

\begin{enumerate}
\item The mean of the time series is constant over time:
\[
E[X_t] = \mu.
\]
\item The covariance and correlation between two observations depend only on the time lag between them rather than on their specific positions in time. These can be expressed as
\[
\gamma_t = \mathrm{Cov}(X_\tau, X_{t+\tau}) = \gamma(\tau)
\]
and
\[
\rho_t = \mathrm{Corr}(X_\tau, X_{t+\tau}) = \rho(\tau),
\]
respectively.
\end{enumerate}

For two such daily series $X_i$ and $X_j$, and a given lag $t$, the cross-covariance is estimated as

\[
c_{ij}(t) = \frac{1}{n} \sum_{s=\max(1,-t)}^{\min(n-t,\,n)} \big[X_i(s+t) - \bar{X}_i\big]\big[X_j(s) - \bar{X}_j\big],
\]

where $\bar{X}_i$ and $\bar{X}_j$ denote the sample means of the respective series for cryptocurrencies $i$ and $j$. The corresponding cross-correlation coefficient at lag $t$ is obtained by normalizing the cross-covariance by the geometric mean of the two series' own variances at lag $0$:

\[
r_{ij}(t) = \frac{c_{ij}(t)}{\sqrt{c_{ii}(0)\, c_{jj}(0)}}.
\]

The cross-correlation coefficient $r_{ij}(t)$ describes the linear relationship between the daily $X$ series of cryptocurrencies $i$ and $j$ at lag $t$. For each cryptocurrency pair, $r_{ij}(t)$ was computed over a range of lags from $t = -10$ to $t = +10$ days, and the lag at which $|r_{ij}(t)|$ attained its maximum value was recorded as the pair's cross-correlation lag (ccf.lag), with the corresponding value of $r_{ij}(t)$ recorded as its cross-correlation coefficient (ccf.cor). Pairs whose overlapping non-missing observations covered less than 20\% of the phase length were excluded from the ranking, and only pairs attaining $|\text{ccf.cor}| > 0.2$ were retained for the results reported in Section~3. All calculations were performed in R, using the \texttt{ccf()} function from the base \texttt{stats} package.

\subsection{Dynamic Time Warping (DTW) Distance}

While the cross-correlation coefficient $r_{ij}(t)$ captures linear co-movement at a fixed integer-day lag, it does not account for cases in which two cryptocurrencies follow a similar price trajectory with a more irregular or variable temporal offset. To complement the cross-correlation analysis, this study additionally computed the DTW distance between each qualifying cryptocurrency pair, using the same daily $X$ series described in Section~2.3 (the signed, capped intraday price-range metric) as the input for both measures.

DTW is a technique for measuring the similarity between two temporal sequences by allowing a nonlinear, elastic alignment (warping) of the time axis, so that sequences with similar shapes are recognized as similar even if they are locally shifted, stretched, or compressed in time \citep{sakoe1978}. For two sequences $X_i = (x_{i,1}, \dots, x_{i,n})$ and $X_j = (x_{j,1}, \dots, x_{j,n})$, DTW finds the warping path $\pi$ that minimizes the cumulative pointwise distance between aligned elements,

\[
\mathrm{DTW}(X_i, X_j) = \min_{\pi} \sum_{(a,b) \in \pi} \left| x_{i,a} - x_{j,b} \right|,
\]

subject to boundary, monotonicity, and continuity constraints on the warping path $\pi$. The resulting distance was normalized by path length to allow comparison across pairs with different degrees of overlap. DTW distances were computed using the \texttt{dtw2vec()} function from the \texttt{IncDTW} package in R, and are reported as \texttt{dtw\_dist} in Section~3, with lower values indicating greater shape similarity between two cryptocurrencies' price dynamics.

DTW distance was computed for every cryptocurrency pair that satisfied the overlap and correlation-based inclusion criteria described in Section~2.3, and is reported alongside the cross-correlation results in Section~3 as a complementary, rather than competing, measure of similarity: whereas $r_{ij}(t)$ identifies pairs whose daily price dynamics move together linearly and contemporaneously, the DTW distance identifies pairs whose overall price trajectories are similar in shape, independent of any single fixed lag.

It should be noted that DTW distance was computed only for pairs that had already exceeded the 
$|\mathrm{ccf.cor}| > 0.2$ threshold described above; consequently, the DTW-based rankings reported in Section~3 are drawn from the same correlation-screened pool of pairs as the cross-correlation rankings, rather than from an independent search over all pairs meeting only the overlap criterion. This means that a pair with very low linear correlation but potentially high shape similarity would not be identified by this procedure, since DTW distance was never computed for it.

\section{Results}
\label{sec:results}

This section presents the results of the cryptocurrency similarity analysis across three distinct market periods. Period 1 corresponds to an upward market trend, Period 2 to a prolonged downward trend, and Period 3 to a subsequent upward trend. For each period, two complementary similarity measures were calculated for cryptocurrency pairs based on the daily $X$ metric described in Section~\ref{sec:methods} (a signed, capped intraday price-range measure): the cross-correlation coefficient (ccf.cor), which captures linear, lag-specific co-movement, and the DTW distance (dtw\_dist), which captures overall shape similarity independent of a fixed lag. The pairs with the highest cross-correlation coefficients are reported first for each period, followed by the pairs with the lowest (most similar) DTW distances, and the two rankings are then compared.

\subsection{Period 1: Upward Trend (March 12, 2020 - November 10, 2021)}

Period 1 was characterized by an overall upward trend in cryptocurrency prices. Cross-correlation coefficients were calculated for all cryptocurrency pairs meeting the inclusion criteria of Section~\ref{sec:methods} (198 cryptocurrencies), using their daily $X$ values. Table \ref{tab:per1} presents the 10 pairs with the highest correlation coefficients.

\begin{table}[ht]
\centering
\begin{tabular}{rllrrr}
  \hline
  & first\_coin & second\_coin & ccf.cor & ccf.lag & n \\ 
  \hline
1 & BNB & CAKE & 0.86 & 0.00 & 265 \\ 
2 & BCH & EOS & 0.85 & 0.00 & 609 \\ 
3 & BCH & LTC & 0.84 & 0.00 & 609 \\ 
4 & NEO & ONT & 0.83 & 0.00 & 609 \\ 
5 & BCH & ETC & 0.81 & 0.00 & 609 \\ 
6 & POLS & ZIL & 0.81 & 0.00 & 176 \\ 
7 & DASH & ZEC & 0.80 & 0.00 & 609 \\ 
8 & NEO & QTUM & 0.79 & 0.00 & 609 \\ 
9 & EOS & ETC & 0.79 & 0.00 & 609 \\ 
10 & ETH & LTC & 0.78 & 0.00 & 609 \\ 
\hline
\end{tabular}
\caption{Top 10 cryptocurrency pairs with the highest cross-correlation coefficients in Period 1 (N = 198). The column $n$ reports the number of overlapping daily observations underlying each pair's estimate, out of a maximum of 609 days for the full phase.}
\label{tab:per1}
\end{table}

The ``first\_coin'' and ``second\_coin'' columns identify the two cryptocurrencies in each pair. The ``ccf.cor'' column reports the cross-correlation coefficient between the two coins' daily $X$ series. The ``ccf.lag'' column reports the time lag, in days, at which the maximum absolute correlation occurred, searched over a window of $\pm 10$ days. As shown in Table \ref{tab:per1}, BNB-CAKE had the highest correlation coefficient (0.86), followed by BCH-EOS (0.85), BCH-LTC (0.84), NEO-ONT (0.83), and BCH-ETC (0.81). All 10 pairs had a lag of 0, indicating that their maximum correlations occurred contemporaneously.

\begin{figure}[htbp]
\centering
\begin{minipage}[t]{0.48\textwidth}
    \centering
    \includegraphics[width=\linewidth]{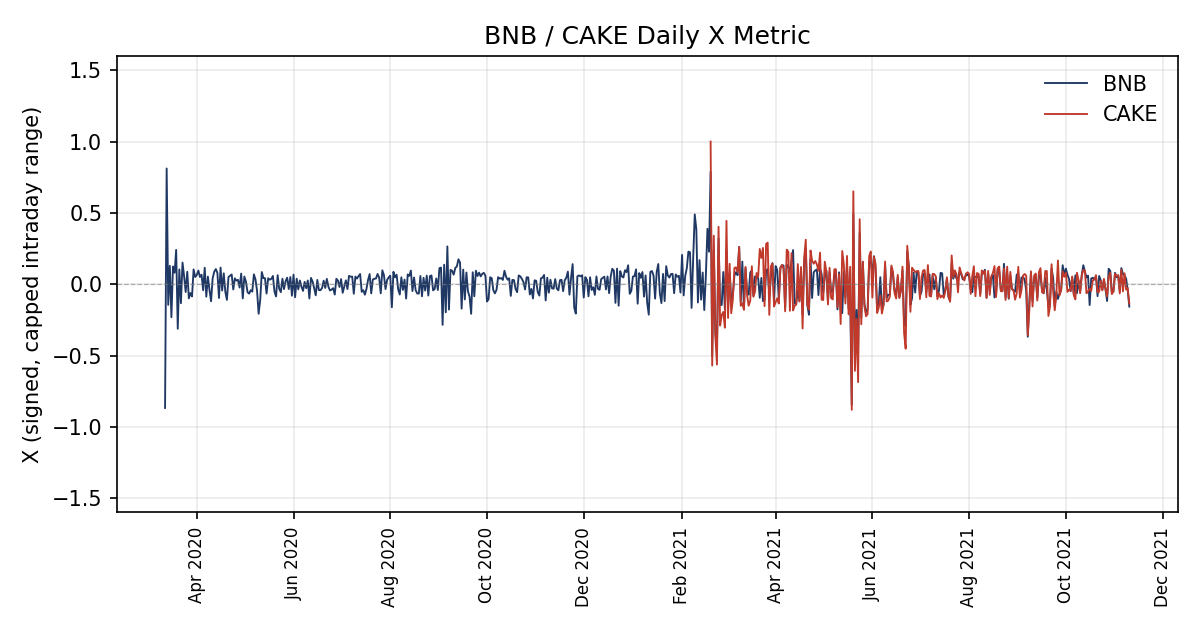}
    \small (a) BNB-CAKE
\end{minipage}
\hfill
\begin{minipage}[t]{0.48\textwidth}
    \centering
    \includegraphics[width=\linewidth]{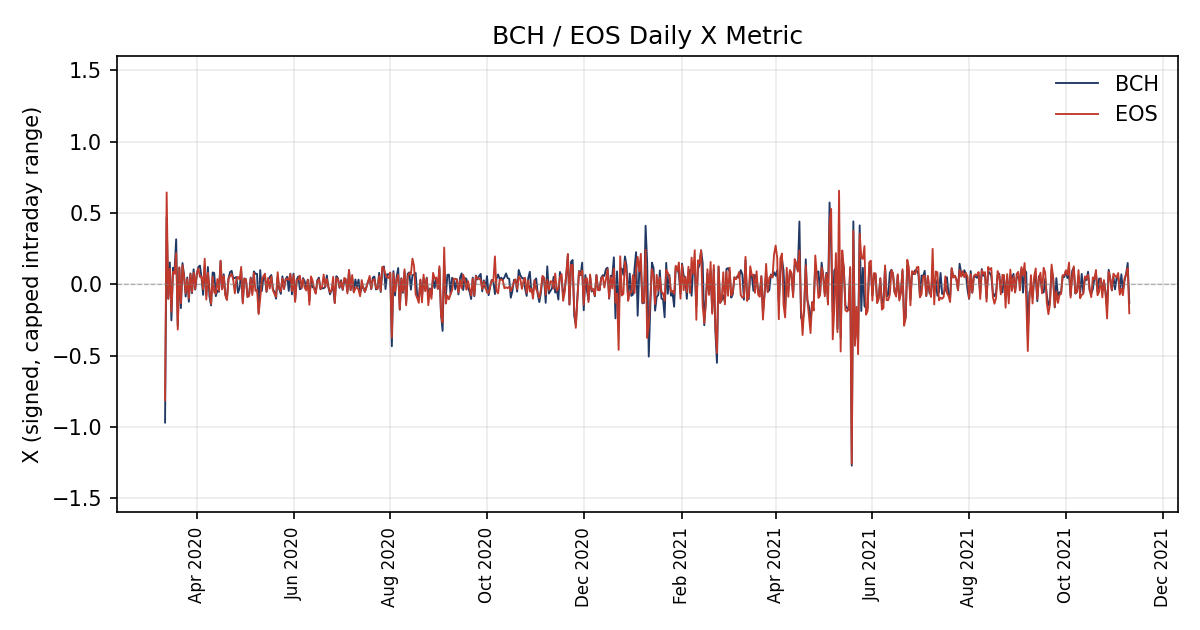}
    \small (b) BCH-EOS
\end{minipage}
\vspace{0.4cm}
\begin{minipage}[t]{0.48\textwidth}
    \centering
    \includegraphics[width=\linewidth]{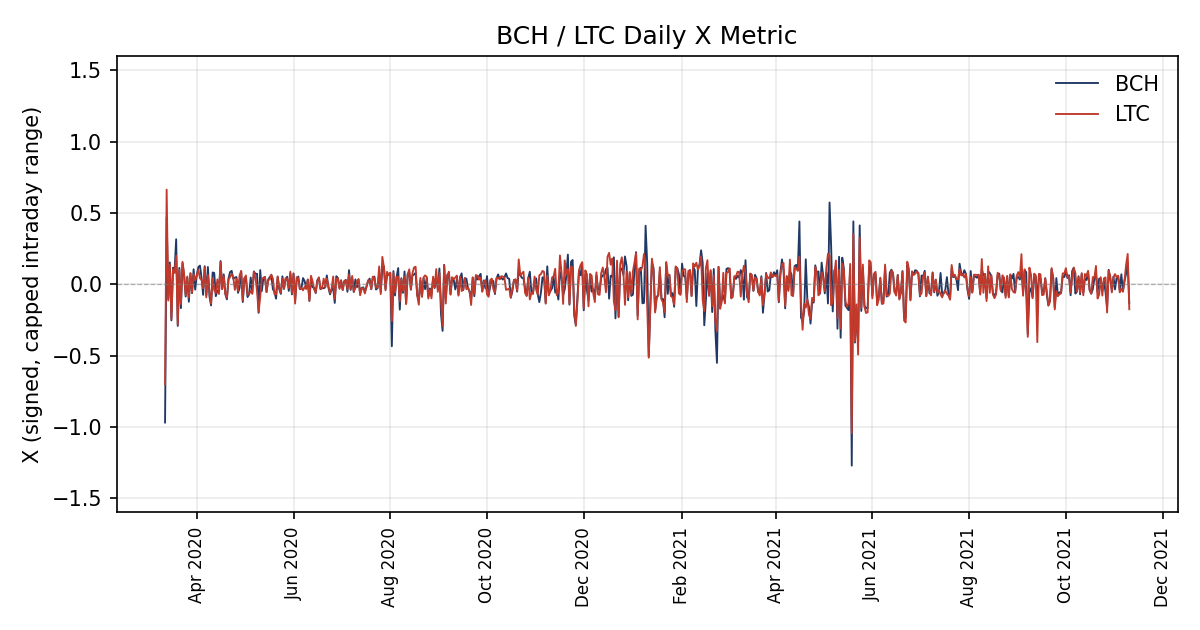}
    \small (c) BCH-LTC
\end{minipage}
\hfill
\begin{minipage}[t]{0.48\textwidth}
    \centering
    \includegraphics[width=\linewidth]{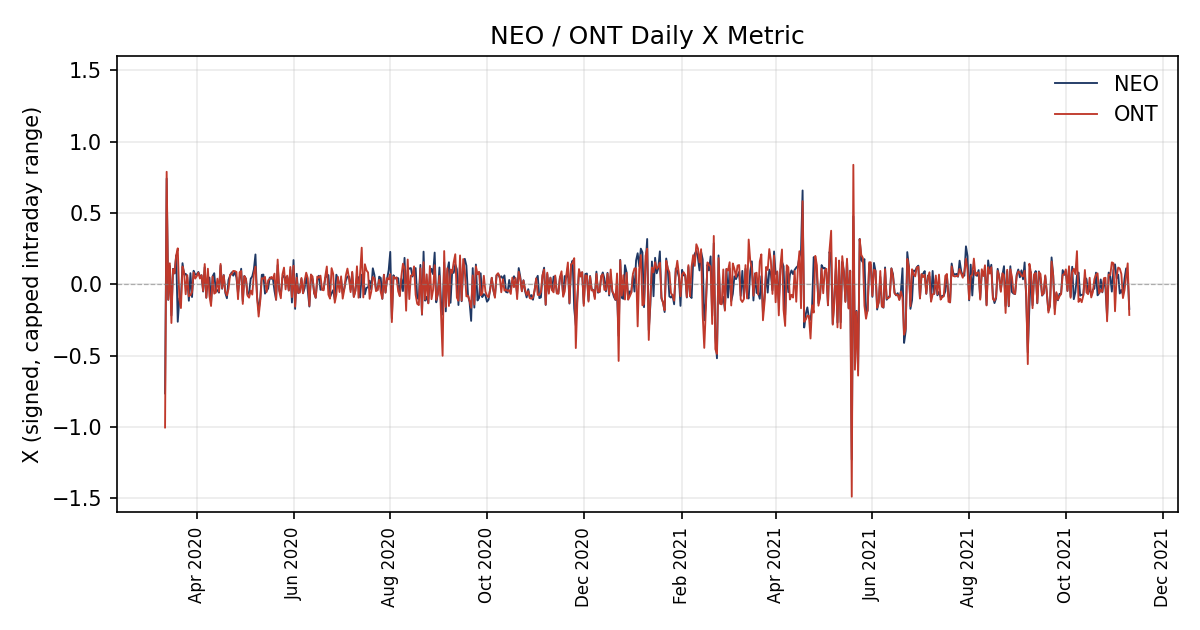}
    \small (d) NEO-ONT
\end{minipage}
\vspace{0.4cm}
\begin{minipage}[t]{0.48\textwidth}
    \centering
    \includegraphics[width=\linewidth]{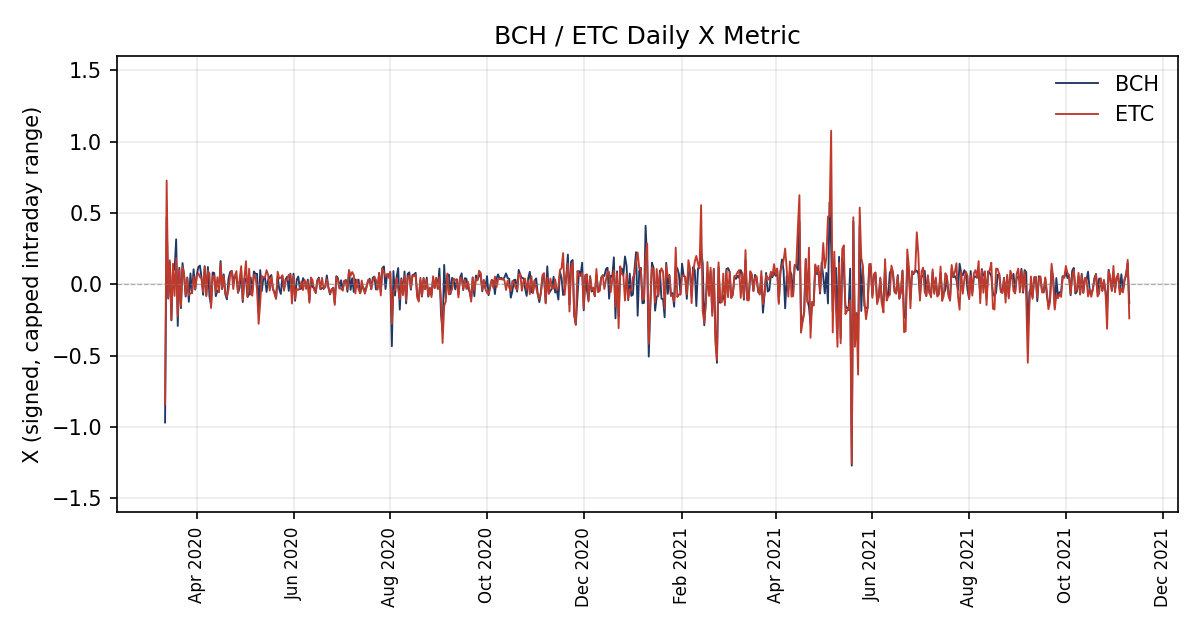}
    \small (e) BCH-ETC
\end{minipage}
\caption{Daily $X$ (signed, capped intraday range) series for the five highest-ranked cryptocurrency pairs in Period 1: (a) BNB-CAKE, (b) BCH-EOS, (c) BCH-LTC, (d) NEO-ONT, and (e) BCH-ETC.}
\label{fig:period1plots}
\end{figure}

Figure \ref{fig:period1plots} displays the daily $X$ series for the five highest-ranked cryptocurrency pairs in Period 1. Because $X$ is bounded and directly comparable across cryptocurrencies, both series in each panel are plotted on the same vertical scale without rescaling. The vertical axis range is held fixed across all panels within a given figure, but differs between figures: it is set according to the largest absolute value of $X$ observed among the pairs shown in that figure, so that each figure remains legible at its own characteristic amplitude rather than being compressed or expanded to match an unrelated figure elsewhere in the manuscript.

Table \ref{tab:dtw1} presents the 10 cryptocurrency pairs with the lowest (most similar) DTW distance in Period 1.

\begin{table}[ht]
\centering
\begin{tabular}{rllrrc}
  \hline
  & first\_coin & second\_coin & dtw.dist & n & in CCF top 10 \\ 
  \hline
1 & EUR & PAXG & 0.008 & 440 &  \\ 
2 & BCH & LTC & 0.025 & 609 & \checkmark \\ 
3 & ETH & LTC & 0.026 & 609 & \checkmark \\ 
4 & BTC & ETH & 0.027 & 609 &  \\ 
5 & NEO & ONT & 0.028 & 609 & \checkmark \\ 
6 & DASH & LTC & 0.029 & 609 &  \\ 
7 & BCH & EOS & 0.029 & 609 & \checkmark \\ 
8 & BCH & ETH & 0.029 & 609 &  \\ 
9 & ETH & FTT & 0.030 & 609 &  \\ 
10 & BTC & FTT & 0.030 & 609 &  \\ 
\hline
\end{tabular}
\caption{Top 10 cryptocurrency pairs with the lowest DTW distance in Period 1. The column $n$ reports the number of overlapping daily observations underlying each pair's estimate.}
\label{tab:dtw1}
\end{table}

Four of the ten lowest-DTW-distance pairs (BCH-LTC, ETH-LTC, NEO-ONT, and BCH-EOS) also appear among the top 10 cross-correlation pairs in Table \ref{tab:per1}, indicating a degree of agreement between the two measures for large, established coins. The remaining pairs, however, do not appear among the top 10 cross-correlation pairs. In particular, EUR-PAXG has the lowest DTW distance of any pair in this period (0.008) despite a comparatively modest cross-correlation coefficient of 0.26, well below the top-10 threshold in Table \ref{tab:per1}. This divergence is addressed further in Section~\ref{sec:dtw-comparison}.

\subsection{Period 2: Downward Trend (November 11, 2021 - October 12, 2023)}

Period 2 was characterized by an overall decline in cryptocurrency prices. Table \ref{tab:per2} presents the 10 cryptocurrency pairs with the highest cross-correlation coefficients during this period, based on 325 cryptocurrencies meeting the inclusion criteria of Section~\ref{sec:methods}.

\begin{table}[htbp]
\centering
\begin{tabular}{rllrrr}
  \hline
  & first\_coin & second\_coin & ccf.cor & ccf.lag & n \\ 
  \hline
1 & XLM & XRP & 0.81 & 0.00 & 701 \\ 
2 & MANA & SAND & 0.81 & 0.00 & 701 \\ 
3 & AGIX & FET & 0.81 & 0.00 & 238 \\ 
4 & BTC & ETH & 0.80 & 0.00 & 701 \\ 
5 & HOT & VET & 0.80 & 0.00 & 701 \\ 
6 & BNB & VET & 0.79 & 0.00 & 701 \\ 
7 & NEO & QTUM & 0.79 & 0.00 & 701 \\ 
8 & ALICE & TLM & 0.79 & 0.00 & 701 \\ 
9 & DOT & VET & 0.79 & 0.00 & 701 \\ 
10 & ONE & VET & 0.79 & 0.00 & 701 \\ 
\hline
\end{tabular}
\caption{Top 10 cryptocurrency pairs with the highest cross-correlation coefficients in Period 2 (N = 325). The column $n$ reports the number of overlapping daily observations underlying each pair's estimate, out of a maximum of 701 days for the full phase.}
\label{tab:per2}
\end{table}

The highest correlation coefficients in Period 2 were observed for XLM-XRP, MANA-SAND, and AGIX-FET, each with a coefficient of 0.81, followed by BTC-ETH and HOT-VET, both at 0.80. The remaining five pairs had coefficients of 0.79. As in Period 1, all 10 pairs had a lag of 0.

\begin{figure}[htbp]
\centering
\begin{minipage}[t]{0.48\textwidth}
    \centering
    \includegraphics[width=\linewidth]{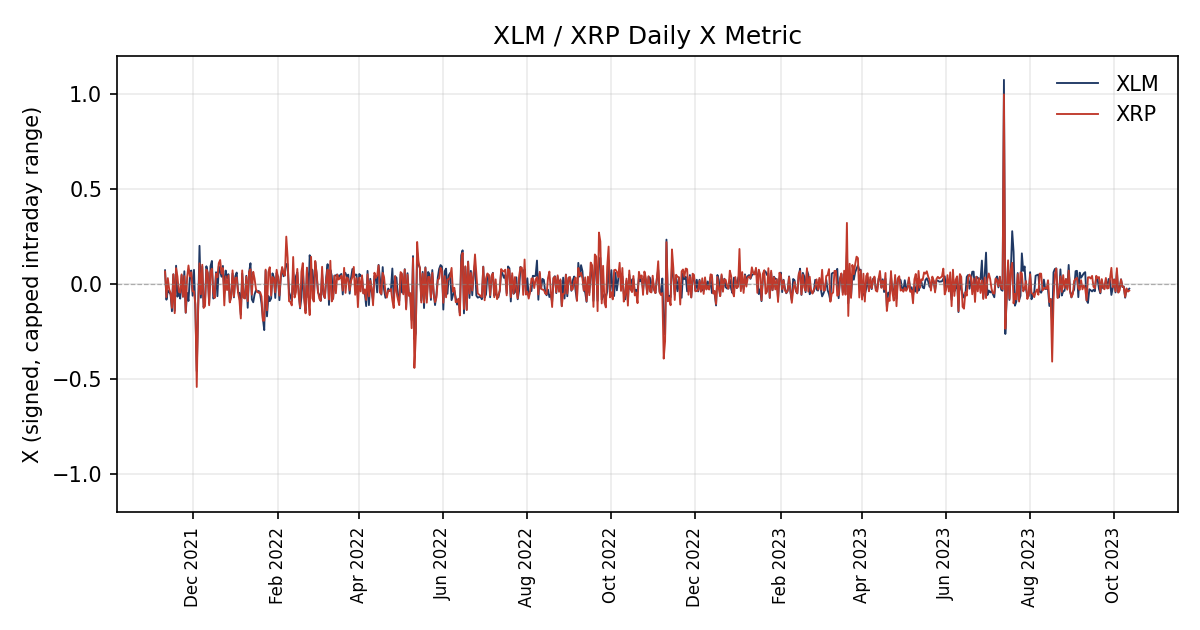}
    \small (a) XLM-XRP
\end{minipage}
\hfill
\begin{minipage}[t]{0.48\textwidth}
    \centering
    \includegraphics[width=\linewidth]{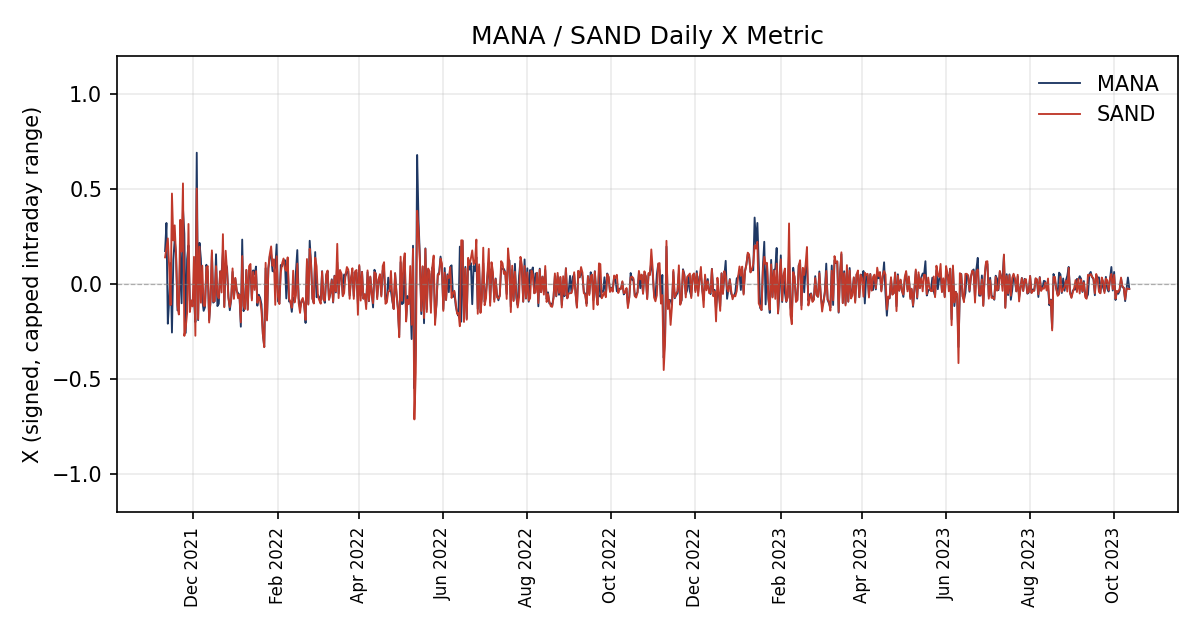}
    \small (b) MANA-SAND
\end{minipage}
\vspace{0.4cm}
\begin{minipage}[t]{0.48\textwidth}
    \centering
    \includegraphics[width=\linewidth]{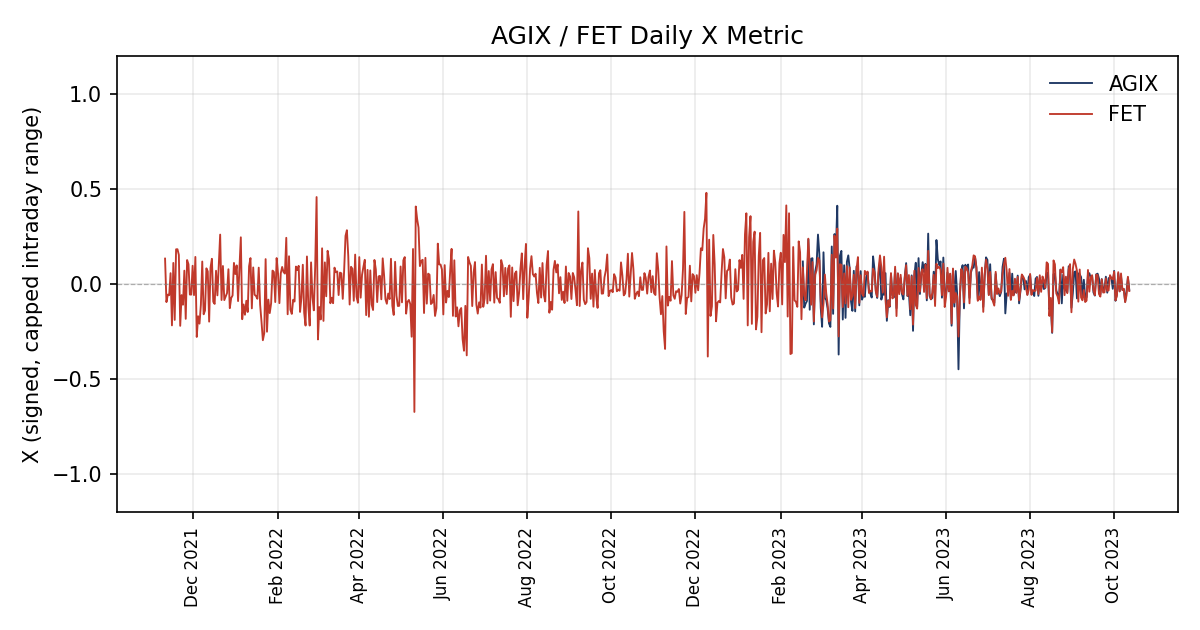}
    \small (c) AGIX-FET
\end{minipage}
\hfill
\begin{minipage}[t]{0.48\textwidth}
    \centering
    \includegraphics[width=\linewidth]{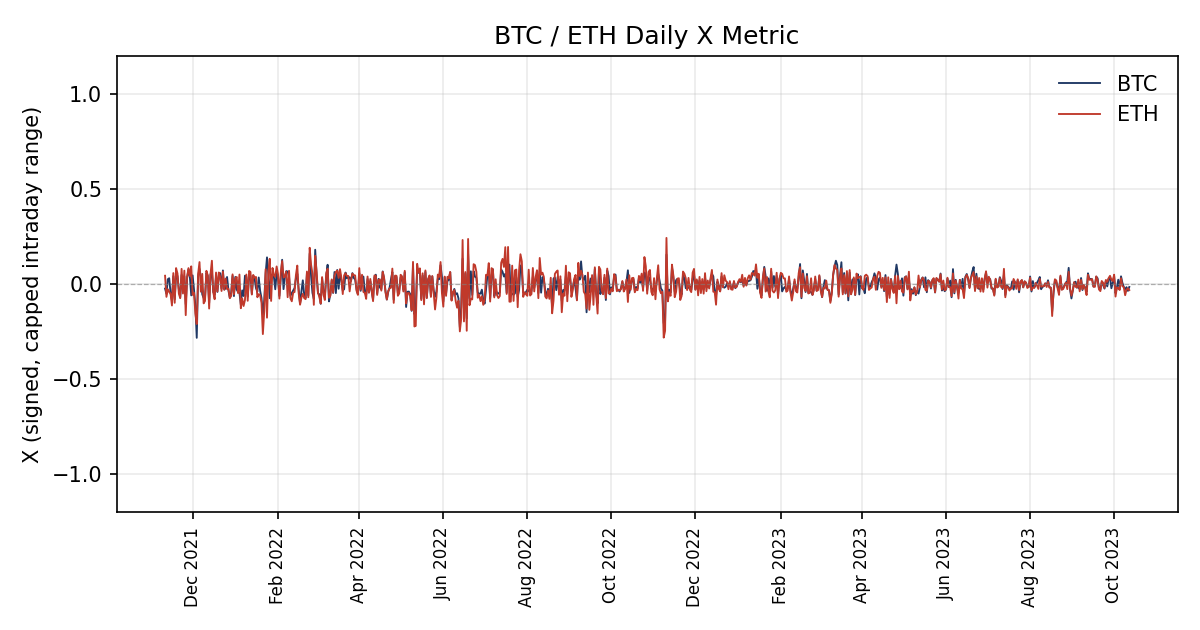}
    \small (d) BTC-ETH
\end{minipage}
\vspace{0.4cm}
\begin{minipage}[t]{0.48\textwidth}
    \centering
    \includegraphics[width=\linewidth]{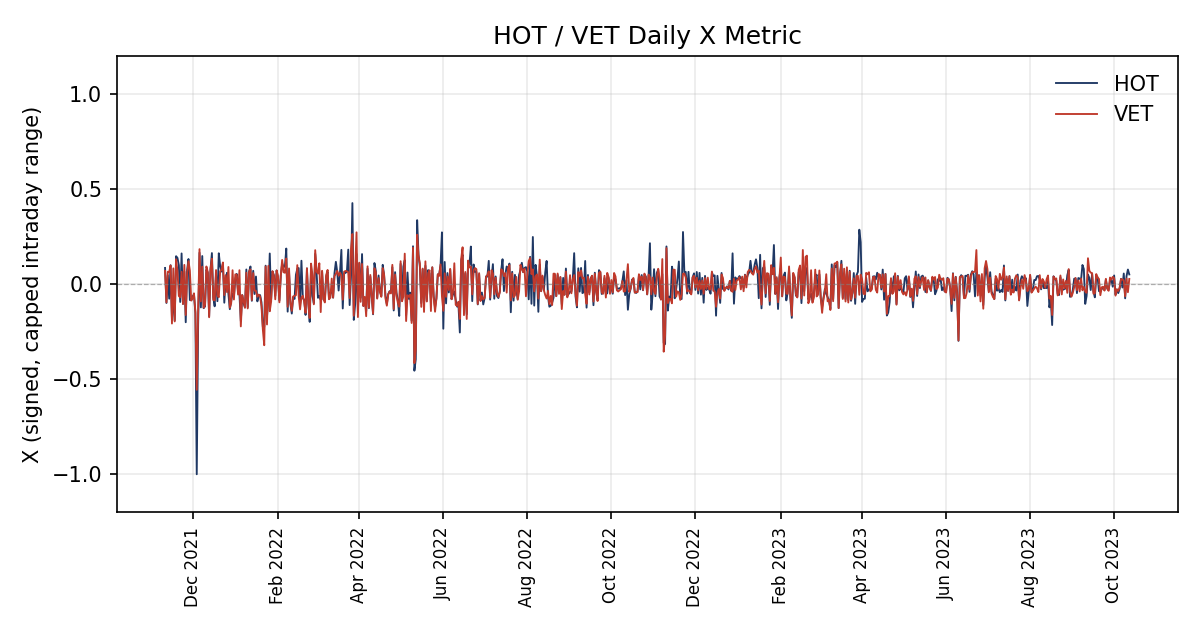}
    \small (e) HOT-VET
\end{minipage}
\caption{Daily $X$ (signed, capped intraday range) series for the five highest-ranked cryptocurrency pairs in Period 2: (a) XLM-XRP, (b) MANA-SAND, (c) AGIX-FET, (d) BTC-ETH, and (e) HOT-VET.}
\label{fig:period2plots}
\end{figure}

Table \ref{tab:dtw2} presents the 10 cryptocurrency pairs with the lowest DTW distance in Period 2.

\begin{table}[ht]
\centering
\begin{tabular}{rllrrc}
  \hline
  & first\_coin & second\_coin & dtw.dist & n & in CCF top 10 \\ 
  \hline
1 & BNB & BTC & 0.016 & 701 &  \\ 
2 & BTC & ETH & 0.017 & 701 & \checkmark \\ 
3 & BNB & CAKE & 0.018 & 701 &  \\ 
4 & BNB & ETH & 0.018 & 701 &  \\ 
5 & BTC & TRX & 0.018 & 701 &  \\ 
6 & BTC & BTTC & 0.018 & 626 &  \\ 
7 & BNB & BTTC & 0.019 & 626 &  \\ 
8 & DOT & VET & 0.019 & 701 & \checkmark \\ 
9 & XLM & XRP & 0.020 & 701 & \checkmark \\ 
10 & ADA & DOT & 0.020 & 701 &  \\ 
\hline
\end{tabular}
\caption{Top 10 cryptocurrency pairs with the lowest DTW distance in Period 2. The column $n$ reports the number of overlapping daily observations underlying each pair's estimate.}
\label{tab:dtw2}
\end{table}

Only three of the ten lowest-DTW-distance pairs in Period 2 (BTC-ETH, DOT-VET, and XLM-XRP) coincide with the cross-correlation top 10. The remaining DTW-ranked pairs are dominated by combinations involving BNB and BTC, suggesting that during this downward-trending phase, the two largest cryptocurrencies by market capitalization exhibited price trajectories that were similar in overall shape, but not necessarily the strongest contemporaneous linear co-movers.

\subsection{Period 3: Upward Trend (October 13, 2023 - April 15, 2024)}

Period 3 represents a subsequent upward trend in cryptocurrency prices. Table \ref{tab:per3} presents the 10 pairs with the highest cross-correlation coefficients, based on 347 cryptocurrencies meeting the inclusion criteria of Section~\ref{sec:methods}.

\begin{table}[htbp]
\centering
\begin{tabular}{rllrrr}
  \hline
  & first\_coin & second\_coin & ccf.cor & ccf.lag & n \\ 
  \hline
1 & MANA & SAND & 0.81 & 0.00 & 186 \\ 
2 & ETH & WNXM & 0.81 & 0.00 & 186 \\ 
3 & DASH & ENJ & 0.80 & 0.00 & 186 \\ 
4 & DASH & SAND & 0.80 & 0.00 & 186 \\ 
5 & ENJ & SAND & 0.80 & 0.00 & 186 \\ 
6 & BTC & ETH & 0.79 & 0.00 & 186 \\ 
7 & XLM & XRP & 0.79 & 0.00 & 186 \\ 
8 & DASH & KNC & 0.78 & 0.00 & 186 \\ 
9 & DASH & ZEC & 0.78 & 0.00 & 186 \\ 
10 & BAT & SXP & 0.78 & 0.00 & 186 \\ 
\hline
\end{tabular}
\caption{Top 10 cryptocurrency pairs with the highest cross-correlation coefficients in Period 3 (N = 347). The column $n$ reports the number of overlapping daily observations underlying each pair's estimate, out of a maximum of 186 days for the full phase.}
\label{tab:per3}
\end{table}

MANA-SAND and ETH-WNXM had the highest correlation coefficients in Period 3, both at 0.81. DASH-ENJ, DASH-SAND, and ENJ-SAND followed at 0.80. BTC-ETH and XLM-XRP had coefficients of 0.79, while DASH-KNC, DASH-ZEC, and BAT-SXP had coefficients of 0.78. All 10 pairs again had a lag of 0. MANA-SAND, XLM-XRP, and BTC-ETH each also appeared among the top 10 pairs in Period 2, indicating that these relationships persisted across the downward and subsequent upward market phases, even as the overall composition and ranking of the most highly correlated pairs varied.

\begin{figure}[htbp]
\centering
\begin{minipage}[t]{0.48\textwidth}
    \centering
    \includegraphics[width=\linewidth]{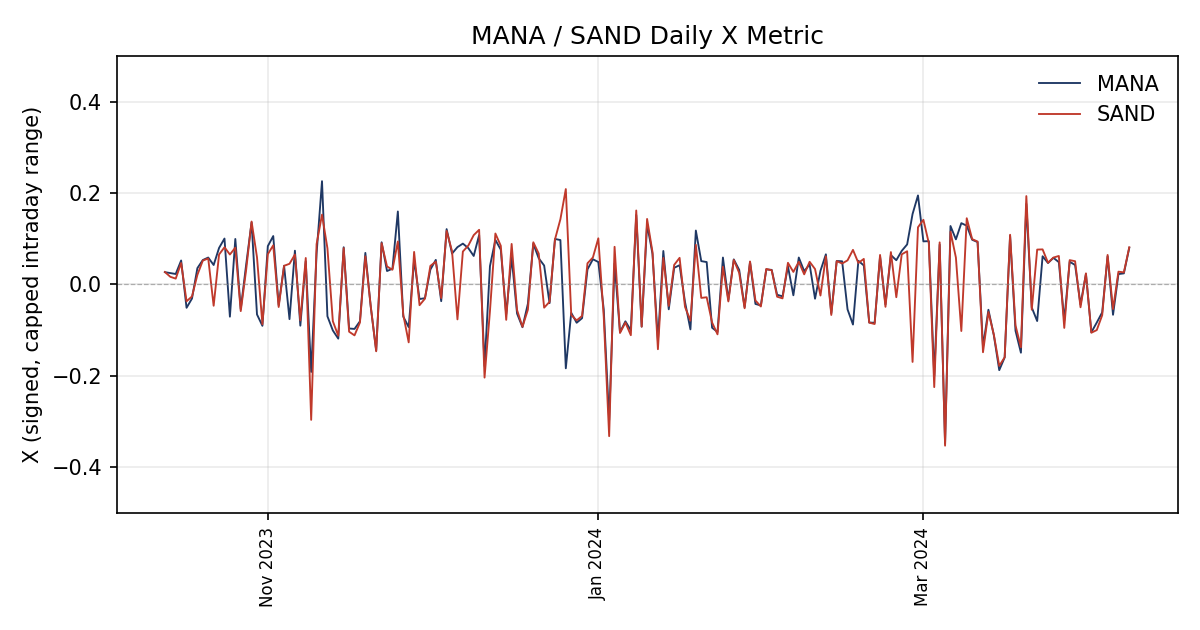}
    \small (a) MANA-SAND
\end{minipage}
\hfill
\begin{minipage}[t]{0.48\textwidth}
    \centering
    \includegraphics[width=\linewidth]{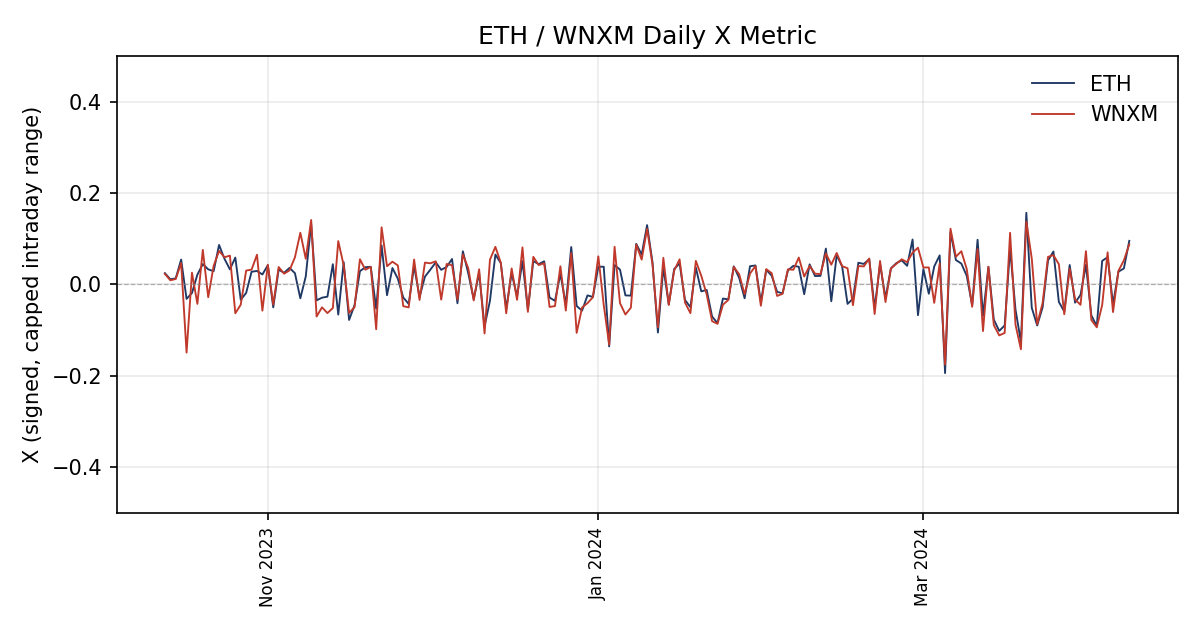}
    \small (b) ETH-WNXM
\end{minipage}
\vspace{0.4cm}
\begin{minipage}[t]{0.48\textwidth}
    \centering
    \includegraphics[width=\linewidth]{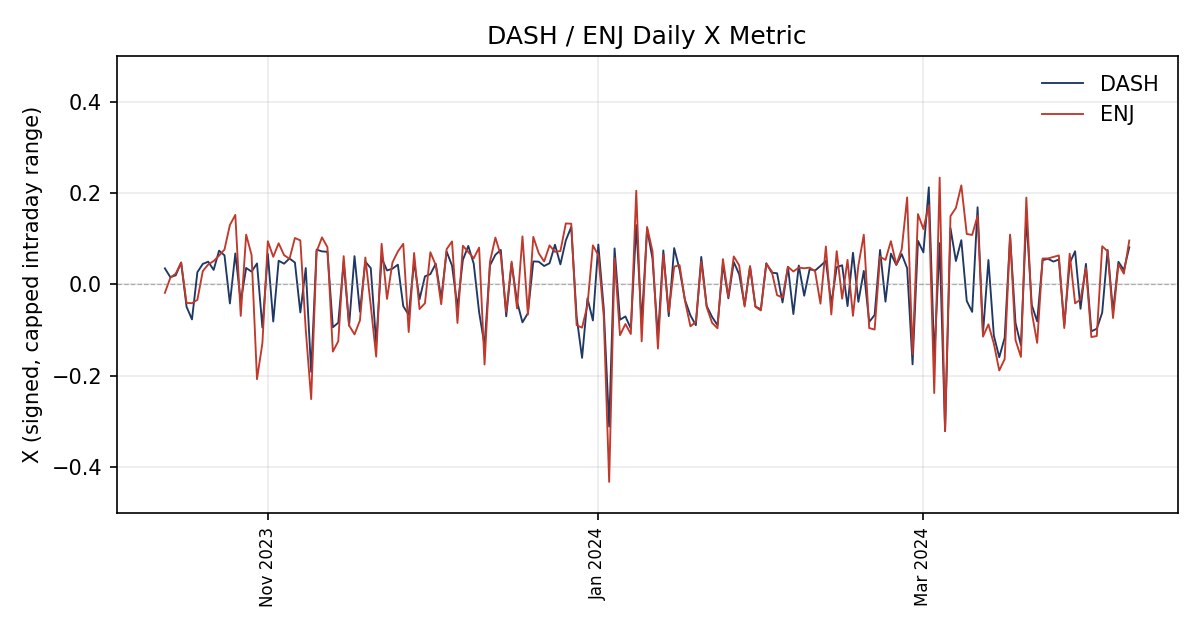}
    \small (c) DASH-ENJ
\end{minipage}
\hfill
\begin{minipage}[t]{0.48\textwidth}
    \centering
    \includegraphics[width=\linewidth]{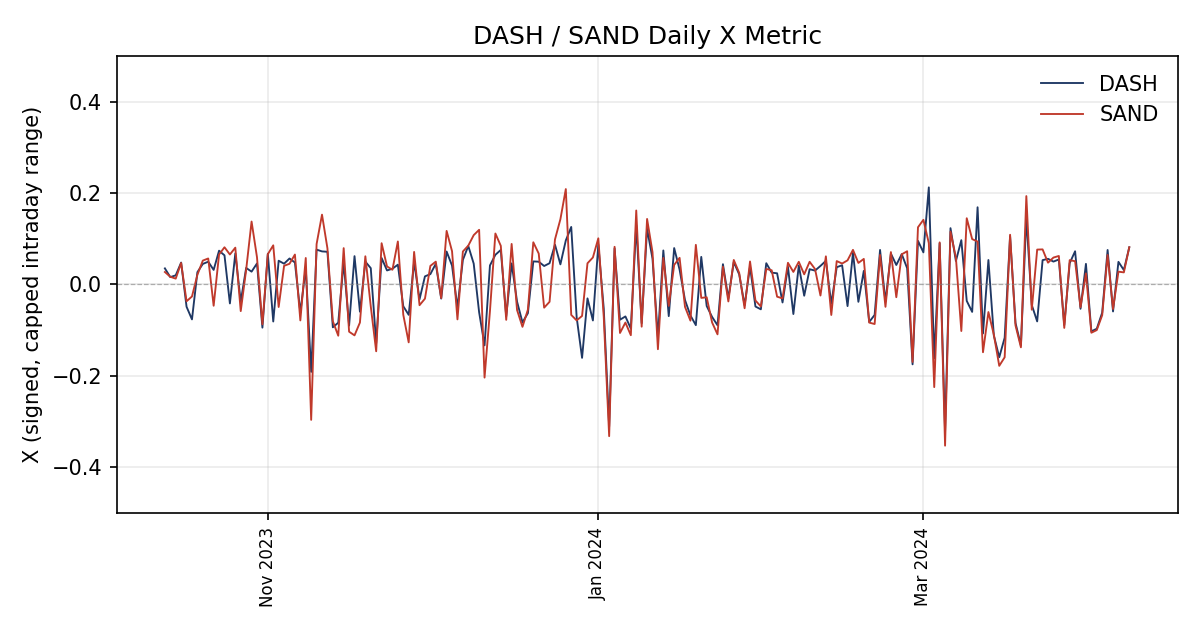}
    \small (d) DASH-SAND
\end{minipage}
\vspace{0.4cm}
\begin{minipage}[t]{0.48\textwidth}
    \centering
    \includegraphics[width=\linewidth]{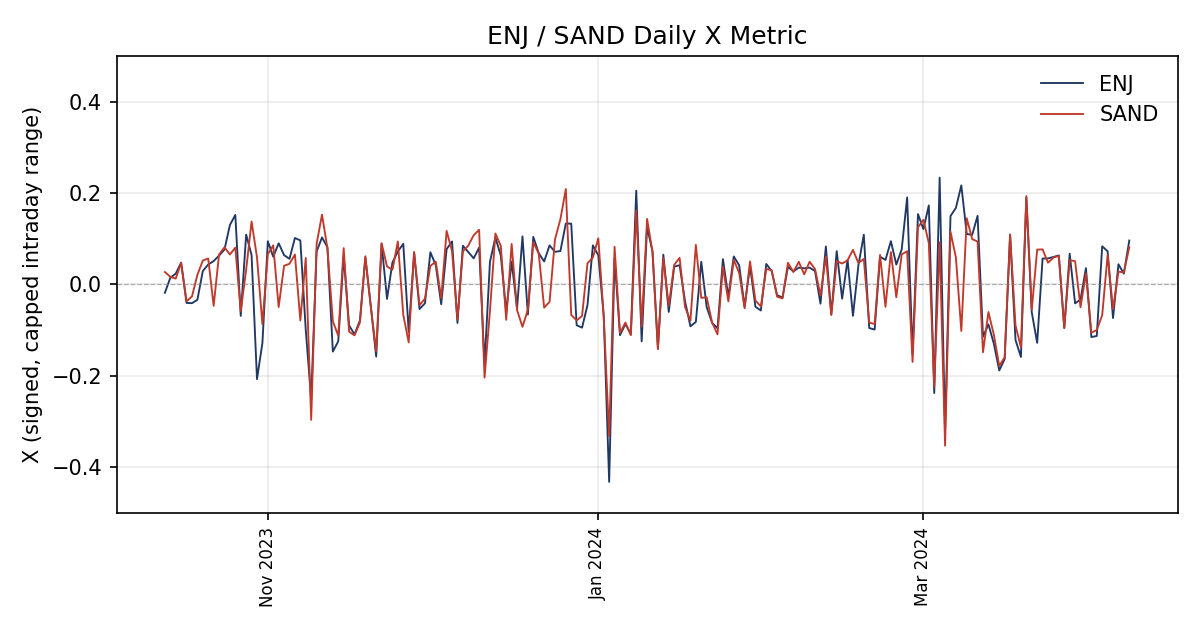}
    \small (e) ENJ-SAND
\end{minipage}
\caption{Daily $X$ (signed, capped intraday range) series for the five highest-ranked cryptocurrency pairs in Period 3: (a) MANA-SAND, (b) ETH-WNXM, (c) DASH-ENJ, (d) DASH-SAND, and (e) ENJ-SAND.}
\label{fig:period3plots}
\end{figure}

Table \ref{tab:dtw3} presents the 10 cryptocurrency pairs with the lowest DTW distance in Period 3.

\begin{table}[ht]
\centering
\begin{tabular}{rllrrc}
  \hline
  & first\_coin & second\_coin & dtw.dist & n & in CCF top 10 \\ 
  \hline
1 & EUR & PAXG & 0.006 & 186 &  \\ 
2 & BTC & ETH & 0.013 & 186 & \checkmark \\ 
3 & ETH & WNXM & 0.015 & 186 & \checkmark \\ 
4 & MANA & SAND & 0.015 & 186 & \checkmark \\ 
5 & ACM & ATM & 0.017 & 186 &  \\ 
6 & ETH & NEXO & 0.017 & 186 &  \\ 
7 & XLM & XRP & 0.017 & 186 & \checkmark \\ 
8 & BTC & TRX & 0.017 & 186 &  \\ 
9 & ACM & CITY & 0.017 & 186 &  \\ 
10 & BNB & ETH & 0.017 & 186 &  \\ 
\hline
\end{tabular}
\caption{Top 10 cryptocurrency pairs with the lowest DTW distance in Period 3. The column $n$ reports the number of overlapping daily observations underlying each pair's estimate.}
\label{tab:dtw3}
\end{table}

As in Period 1, EUR-PAXG again has the lowest DTW distance of any pair (0.006), despite a modest cross-correlation coefficient of 0.23. Notably, two fan-token pairs issued on the same tokenization platform, ACM-ATM and ACM-CITY, also appear among the ten lowest-DTW-distance pairs in this period, despite neither pair reaching the cross-correlation top 10. Unlike EUR-PAXG, these fan tokens exhibit daily volatility comparable to BTC and ETH over this period (standard deviation of $X$ of approximately 0.06–0.07 for all three tokens, versus 0.05–0.07 for BTC and ETH), indicating that their DTW-based similarity does not simply reflect low-amplitude, near-flat price dynamics, but rather a genuine shared shape in their day-to-day price trajectories, plausibly related to common platform-level news or liquidity events affecting fan tokens as a category.

\subsection{Comparison of Cross-Correlation and DTW Rankings}
\label{sec:dtw-comparison}

Across the three market phases, between three and four of the ten lowest-DTW-distance pairs also appeared among the ten highest cross-correlation pairs (four in Period 1, three in Period 2, and four in Period 3). This partial overlap indicates that the two measures are related but not interchangeable: pairs identified as highly similar by both measures, such as BCH-LTC and BCH-EOS in Period 1, BTC-ETH in Period 2, and MANA-SAND and XLM-XRP in Period 3, represent cases of price dynamics that are simultaneously strong linear co-movers and similar in overall shape.

\begin{figure}[htbp]
\centering
\begin{minipage}[t]{0.48\textwidth}
    \centering
    \includegraphics[width=\linewidth]{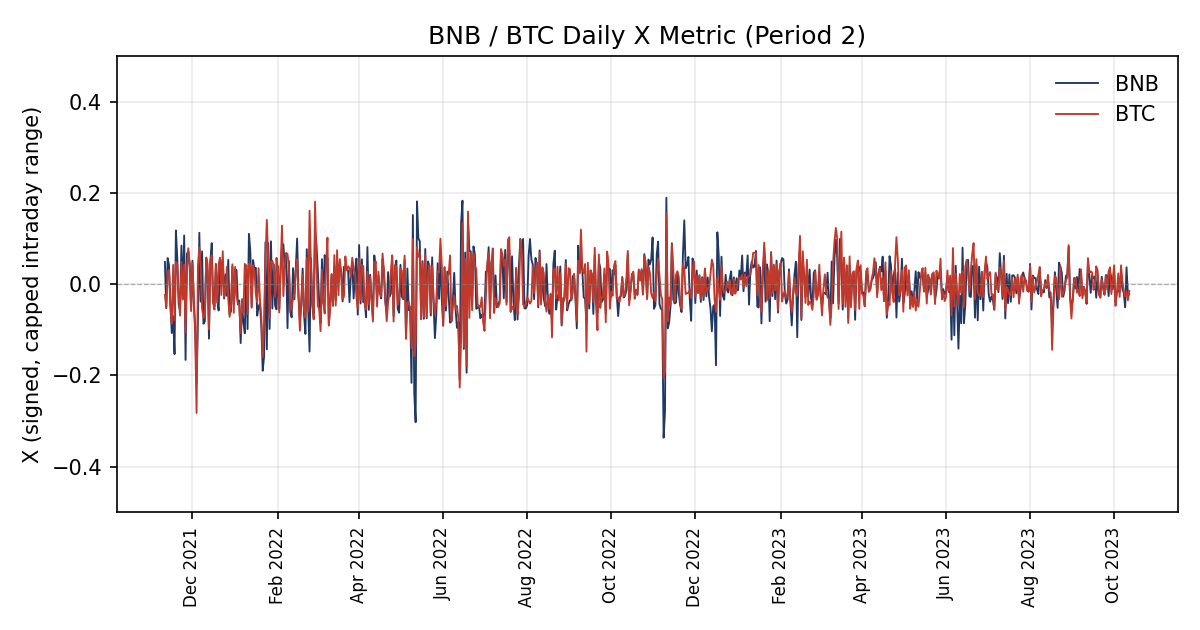}
    \small (a) BNB-BTC (Period 2)
\end{minipage}
\hfill
\begin{minipage}[t]{0.48\textwidth}
    \centering
    \includegraphics[width=\linewidth]{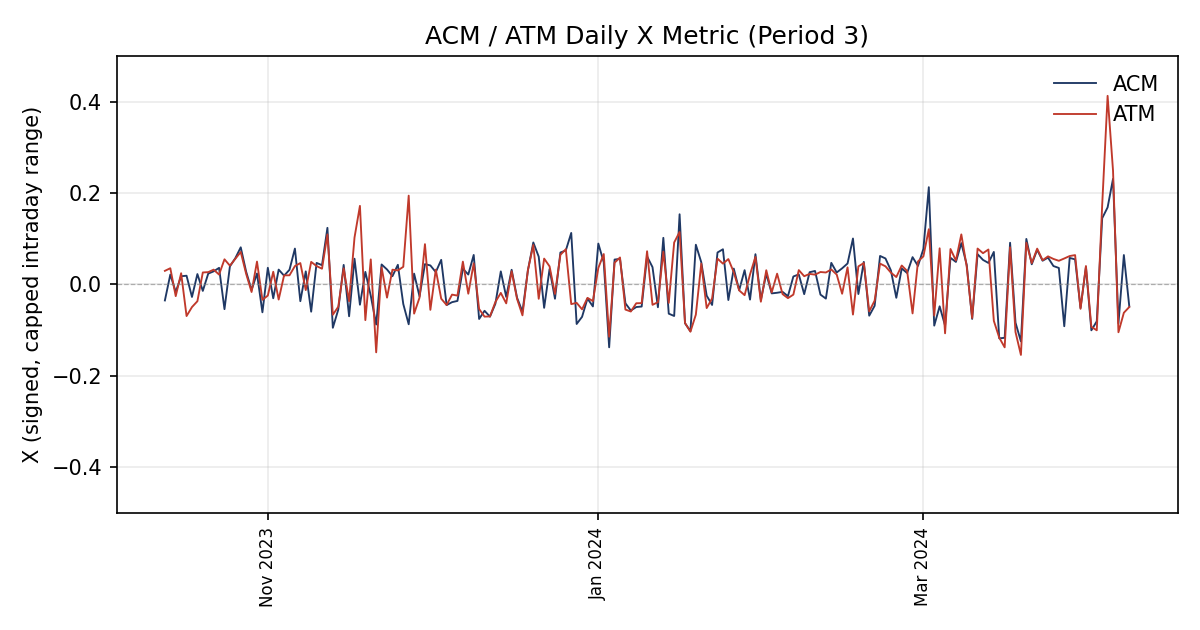}
    \small (b) ACM-ATM (Period 3)
\end{minipage}
\vspace{0.4cm}
\begin{minipage}[t]{0.48\textwidth}
    \centering
    \includegraphics[width=\linewidth]{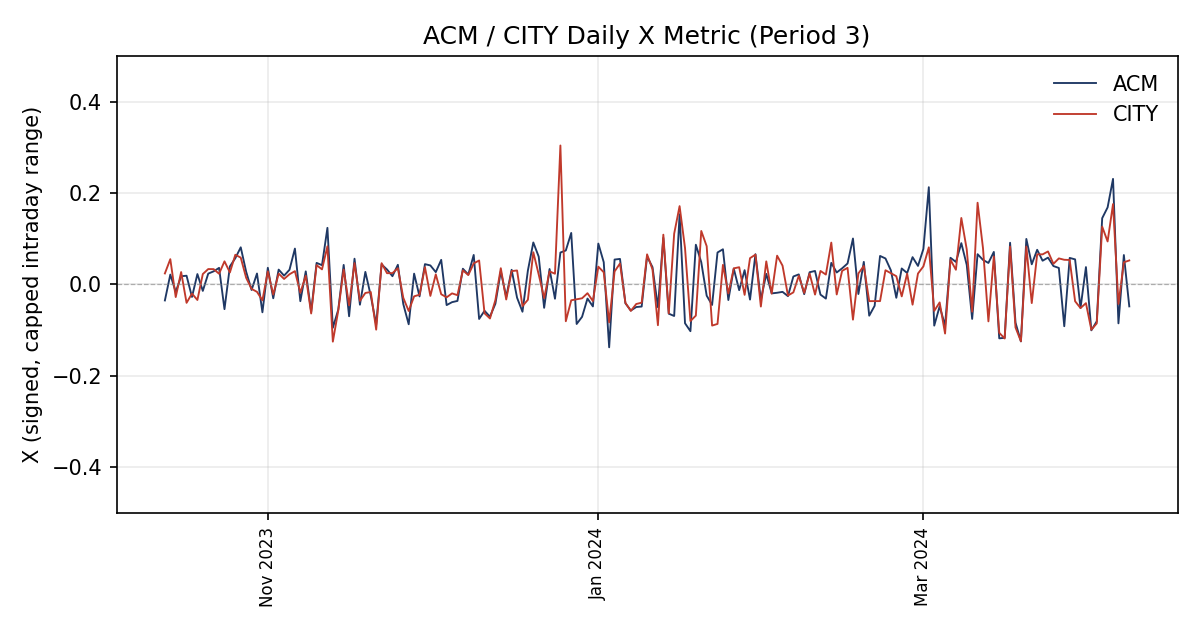}
    \small (c) ACM-CITY (Period 3)
\end{minipage}
\caption{Daily $X$ series for three cryptocurrency pairs identified as highly similar by DTW distance but not by cross-correlation: (a) BNB-BTC in Period 2, (b) ACM-ATM in Period 3, and (c) ACM-CITY in Period 3.}
\label{fig:dtw_genuine}
\end{figure}

Pairs identified by DTW alone point to a different, complementary kind of relationship. Figure \ref{fig:dtw_genuine} displays the daily $X$ series for three such pairs: BNB-BTC in Period 2, and ACM-ATM and ACM-CITY, two fan tokens issued on the same tokenization platform, in Period 3. In all three cases, the two series exhibit clearly visible co-movement in the magnitude and timing of their daily swings, despite none of these pairs reaching the cross-correlation top 10 for their respective period. This suggests that, among pairs with at least a weak-to-moderate linear correlation, DTW is capturing genuine similarity in overall price dynamics that a fixed-lag linear measure does not fully register, plausibly reflecting shared exposure to platform-level or market-cap-tier-specific news and liquidity events.

\begin{figure}[htbp]
\centering
\begin{minipage}[t]{0.48\textwidth}
    \centering
    \includegraphics[width=\linewidth]{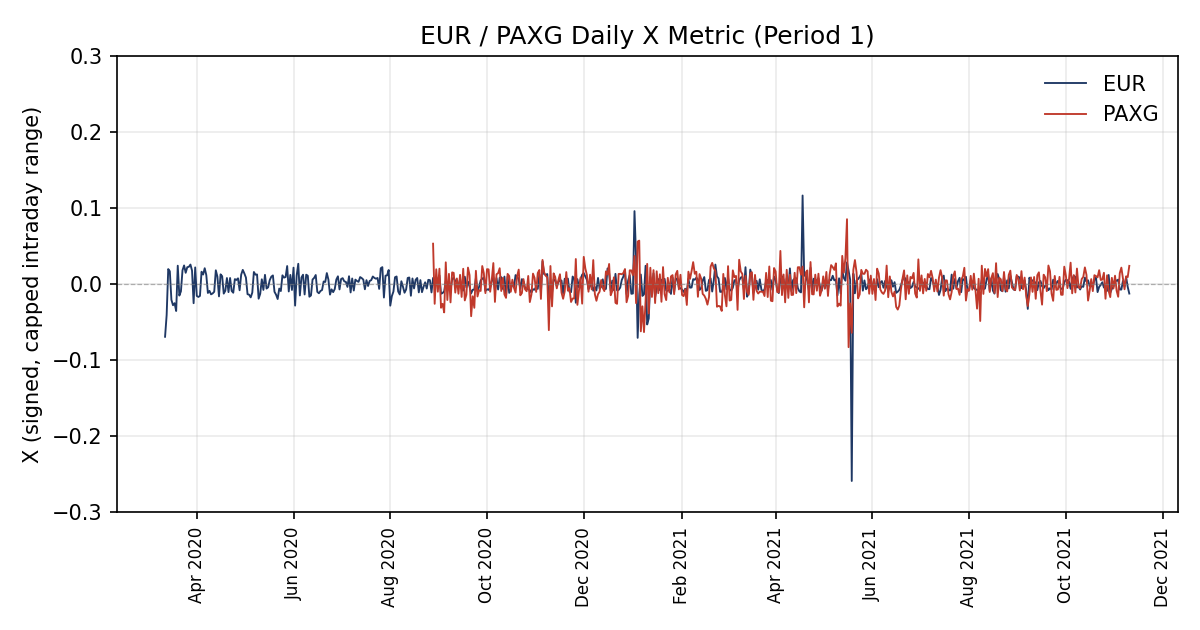}
    \small (a) Period 1
\end{minipage}
\hfill
\begin{minipage}[t]{0.48\textwidth}
    \centering
    \includegraphics[width=\linewidth]{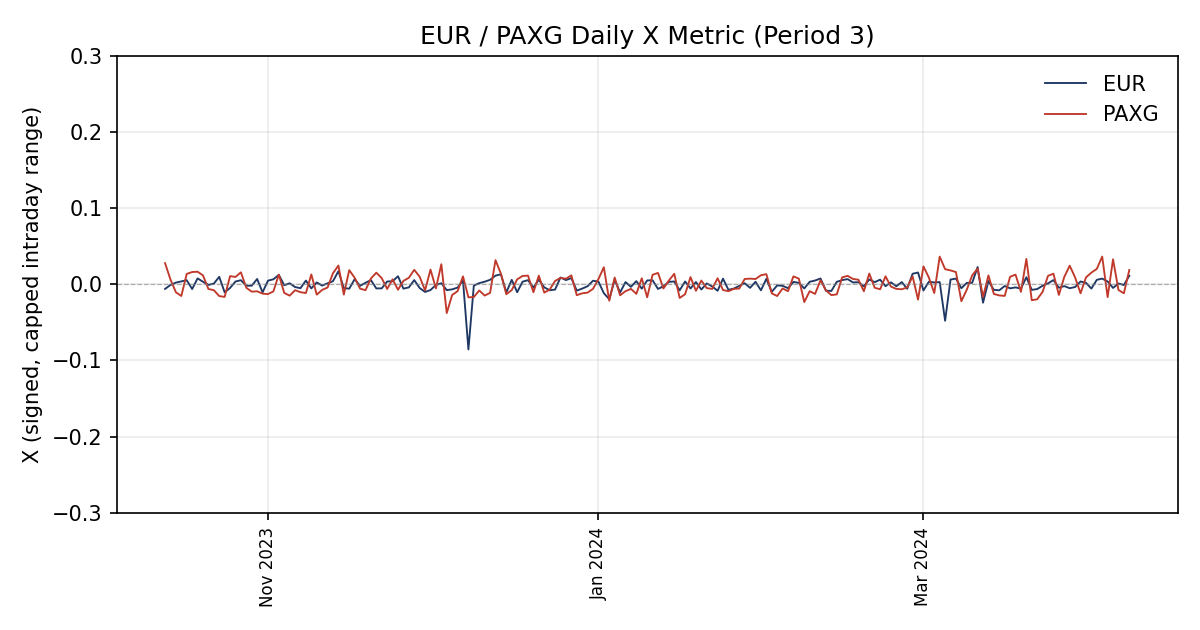}
    \small (b) Period 3
\end{minipage}
\caption{Daily $X$ series for EUR-PAXG, the pair with the lowest DTW distance in both (a) Period 1 and (b) Period 3. Both series remain close to zero for most of the period, illustrating how low DTW distance can arise from uniformly low volatility rather than from strong shared dynamics.}
\label{fig:eurpaxg}
\end{figure}

One limitation of the DTW-based ranking should be noted, however. In both Period 1 and Period 3, the pair with the single lowest DTW distance, EUR-PAXG, exhibited a comparatively low cross-correlation coefficient (0.26 and 0.23, respectively) and substantially lower volatility of $X$ than the other top-ranked pairs. Figure \ref{fig:eurpaxg} displays the daily $X$ series for this pair in both periods. In contrast to the pairs in Figure \ref{fig:dtw_genuine}, both series here remain close to zero for almost the entire period, with only occasional small spikes. Because DTW distance reflects the cumulative pointwise distance along the optimal alignment path, two series with uniformly small day-to-day fluctuations can achieve a low DTW distance even when their fluctuations are not strongly related to one another. Consequently, low DTW distance is most reliably interpreted as evidence of genuine shape similarity when it co-occurs with a non-trivial cross-correlation coefficient and comparable volatility between the two series, as in Figure \ref{fig:dtw_genuine}, rather than in isolation.

\section{Discussion and Conclusion}
\label{sec:conclusion}

This study examined price relationships among 347 cryptocurrencies across three distinct market phases over the period from 2020 to 2024, using two complementary similarity measures: Pearson cross-correlation, based on a daily signed, capped intraday price-range metric ($X$), and Dynamic Time Warping (DTW) distance. Cross-correlation revealed strong positive co-movement among the highest-ranked cryptocurrency pairs in each market phase, with coefficients of 0.78 or higher for the 10 leading pairs reported in each period. BNB-CAKE exhibited the highest correlation during the first upward-trend period, with a coefficient of 0.86. During the downward-trend period, XLM-XRP, MANA-SAND, and AGIX-FET had the highest coefficients, each at 0.81, while MANA-SAND and ETH-WNXM had the highest correlations in the subsequent upward-trend period, both at 0.81.

The composition and ranking of the most highly correlated pairs varied across the three market phases. Nevertheless, some pairs appeared among the highest-ranked pairs in more than one period. MANA-SAND had a correlation coefficient of 0.81 in both Periods 2 and 3, while XLM-XRP had coefficients of 0.81 and 0.79, respectively. BTC-ETH also appeared among the top 10 pairs in both periods, with coefficients of 0.80 and 0.79. These findings indicate that some strong pairwise relationships persisted across the downward and subsequent upward market phases, while other highly correlated pairs differed across periods. This variation is consistent with previous research showing that dependence and connectedness among cryptocurrencies can change over time and across market conditions \citep{ANTONAKAKIS201937, CORBET201828}. From a portfolio perspective, the presence of strongly positively correlated cryptocurrency pairs suggests that holding such assets together may provide limited diversification benefits. At the same time, changes in the identities and rankings of the most highly correlated pairs across market phases highlight the importance of considering market-dependent co-movement when evaluating cryptocurrency diversification strategies.

For all of the top 10 cryptocurrency pairs reported in each of the three periods, the maximum cross-correlation occurred at lag 0. Thus, the strongest relationships among these selected pairs were contemporaneous rather than characterized by observable lead-lag patterns. This result should not, however, be interpreted as evidence of market efficiency or rapid information transmission, since the present analysis was designed to characterize pairwise price co-movement rather than to test these mechanisms.

The DTW analysis complemented these findings by identifying pairs whose overall price trajectories were similar in shape, independent of a fixed lag. Between three and four of the ten lowest-DTW-distance pairs coincided with the ten highest cross-correlation pairs in each period, indicating that the two measures are related but capture distinct aspects of similarity. Two results from the DTW analysis are worth highlighting. First, in Period 3, two fan tokens issued on the same tokenization platform, ACM-ATM and ACM-CITY, ranked among the lowest-DTW-distance pairs despite not reaching the cross-correlation top 10, and exhibited daily volatility comparable to BTC and ETH rather than unusually flat price dynamics. This suggests that, among pairs with at least a weak-to-moderate linear correlation, DTW can surface economically meaningful co-movement, plausibly linked to shared platform-level exposure, that a fixed-lag linear measure does not fully register. Second, the pair with the single lowest DTW distance in both Period 1 and Period 3, EUR-PAXG, exhibited comparatively low cross-correlation and substantially lower volatility than the other top-ranked pairs, illustrating that a low DTW distance can also arise from two series that are both unusually flat rather than from genuine shared dynamics. Taken together, these results suggest that DTW distance is most informative when interpreted alongside cross-correlation and volatility, rather than in isolation, and that combining a linear, lag-specific measure with a shape-based, lag-flexible measure provides a more complete picture of cryptocurrency co-movement than either would in isolation.

Several limitations should be considered when interpreting the findings. First, cross-correlation captures linear, contemporaneous or fixed-lag relationships, and while DTW relaxes the fixed-lag assumption, neither measure was subjected to formal statistical inference: no confidence intervals were constructed and no correction for multiple comparisons was applied when identifying the highest-ranked pairs among the very large number of pairwise combinations evaluated in each period. The reported relationships should therefore be interpreted as descriptive rather than as formal tests of statistical significance, and should not be taken as evidence of causal relationships between cryptocurrencies. Second, pairs were admitted to the ranking procedure if their overlapping non-missing observations covered at least 20\% of the phase length, which for Period~1 corresponds to as few as approximately 122 days; estimates based on shorter overlaps are subject to greater sampling variability than those based on the full phase, and the ranking procedure does not explicitly penalize shorter overlaps. As reported in Tables~\ref{tab:per1}--\ref{tab:dtw3}, the great majority of the highest-ranked pairs are based on the full phase length, with the shortest overlap among the reported cross-correlation pairs corresponding to 176 days in Period~1 (29\% of the phase) and 238 days in Period~2 (34\% of the phase); nonetheless, this heterogeneity in the number of underlying observations should be kept in mind when comparing coefficients across pairs. Third, the DTW-based rankings reported in Section~3 are not the result of an independent search over all cryptocurrency pairs: as noted in Section~2.4, DTW distance was computed only for pairs whose cross-correlation coefficient already exceeded the $|\text{ccf.cor}| > 0.2$ threshold. Consequently, the reported complementarity between cross-correlation and DTW should be understood as arising within this correlation-screened pool of pairs, rather than as evidence that DTW can identify pairs with strong shape similarity but negligible linear correlation; such pairs, if they exist, were not evaluated by the DTW procedure used here. Fourth, the three market phases were defined according to broad market trends rather than estimated using a formal statistical regime-identification procedure. Fifth, the dataset was restricted to cryptocurrencies available through Binance and expressed in USDT, which may limit the generalizability of the results to cryptocurrencies traded on other exchanges or against other currencies; missing observations and the availability of data for cryptocurrencies that were delisted or not available throughout the study period also represent limitations of the dataset. Finally, both the cross-correlation and DTW analyses focus on the pairs with the strongest observed similarity in each period, rather than characterizing the full distribution of pairwise relationships across the broader set of cryptocurrencies.

Future research could extend the present analysis by incorporating formal inferential procedures, such as significance testing with correction for multiple comparisons, statistically estimated market regimes, and data from additional cryptocurrency exchanges and trading pairs. Further studies could also investigate the potential roles of macroeconomic conditions, regulatory developments, technological characteristics, and other market-specific factors, including shared platform-level exposure of the kind observed for fan tokens in this study, in explaining changes in cryptocurrency dependence structures. A natural extension of the DTW analysis in particular would be to relax the correlation-based pre-screening described above, so as to search for shape similarity independently of linear correlation across the full set of eligible pairs. Such extensions could provide a more comprehensive understanding of how relationships among cryptocurrencies evolve over time.

Overall, the findings demonstrate that strong co-movement is present among selected cryptocurrency pairs under different market conditions, whether measured through linear cross-correlation or shape-based DTW distance, but that the identities and relative rankings of the most similar pairs are not constant across market phases and depend in part on which measure of similarity is used. The predominance of lag-0 relationships among the highest cross-correlation pairs indicates that their strongest observed linear associations were contemporaneous, while the partial disagreement between cross-correlation and DTW rankings shows that these two measures are complementary rather than interchangeable within the correlation-screened pool of pairs considered here. Taken together, these results emphasize the dynamic and multidimensional nature of cryptocurrency co-movement and the importance of accounting for both changing correlation structures and the choice of similarity measure when evaluating relationships and diversification opportunities within cryptocurrency markets.

\section*{Data and Code Availability}
The data and R code used to conduct the analyses and generate the figures reported in this study are available at \url{https://github.com/onurbdogan/crypto-ccf-dtw-analysis}.

\section*{Funding}

No funding was received for this research.

\section*{Disclosure statement}

The authors report there are no competing interests to declare.

\addcontentsline{toc}{section}{\refname}
\bibliographystyle{sa-ijas}  

\bibliography{sample}

\end{document}